\documentclass[12pt,titlepage]{article}
\usepackage{amsmath,amsfonts,graphicx}

\title{Hamiltonian cycles on random lattices of arbitrary genus}

\author{%
  Saburo Higuchi\thanks{e-mail: hig@rice.c.u-tokyo.ac.jp}\\
  {\it Department of Pure and Applied Sciences,} \\
  {\it The University of Tokyo} \\
  {\it  Komaba, Meguro, Tokyo 153-8902, Japan}
  }

\renewcommand{\thefootnote}{\fnsymbol{footnote}}

\begin{document}
%\maketitle
\begin{titlepage}
\thispagestyle{empty}
  \begin{flushright}
    June 1998\\
(revised) October 1998\\
    UT-KOMABA/98-16\\
    cond-mat/9806349
  \end{flushright}
\vspace*{\fill}

\begin{center}
{\LARGE\bf   Hamiltonian cycles }\\[0.4ex]  

{\LARGE\bf \mbox{on random lattices of arbitrary genus}}
\vspace*{\fill}

{\Large Saburo Higuchi}%
\footnote{e-mail: \texttt{hig@rice.c.u-tokyo.ac.jp}}\\[0.5ex]

{\it Department of Pure and Applied Sciences,} \\
{\it The University of Tokyo} \\
{\it  Komaba, Meguro, Tokyo 153-8902, Japan}
  \end{center}
\vspace*{\fill}

\vspace*{\fill}
%\newpage
\begin{center}
{\large\bf Abstract}
\end{center}
\smallskip 
A Hamiltonian cycle of a graph is a closed path that visits every
vertex once and only once. It has been difficult to count the number
of Hamiltonian cycles on regular lattices with periodic boundary
conditions, \textsl{e.g.} lattices on a torus, due to the presence of
winding modes.  In this paper, the exact number of Hamiltonian cycles
on a random trivalent fat graph drawn faithfully on a torus is
obtained.  This result is further extended to the case of random
graphs drawn on surfaces of an arbitrary genus.  The conformational
exponent $\gamma$ is found to depend on the genus linearly.
\vspace*{5ex}

\noindent
PACS:
05.20.-y; %Statistical mechanics
02.10.Eb; %Combinatorics
04.60.Nc; % Lattice and discrete method
82.35.+t  %Polymer reactions and polymerization

\noindent
Keywords: 
random graph; 
random lattice; 
Hamiltonian cycle;
self-avoiding walk;
compact polymer
\end{titlepage}

\setcounter{footnote}{0}
\renewcommand{\thefootnote}{\arabic{footnote}}

\section{Introduction}
Properties of Hamiltonian cycles and walks on regular and random
lattices have been attracting much attention recently 
\cite{JaKo:fieldtheory,cond-mat/9711152,KoJa:conformational,%
KodeNi:FPLhoneycomb,EyGuKr:hamiltonian,Higuchi:hamiltonianrandom}.
A Hamiltonian cycle (walk) of a graph is a closed (open) path
which visits every vertex once and only once, \text{i.e.},
a self-avoiding loop (walk) which visits all the vertices.
The system of a Hamiltonian cycle (walk) on a lattice serves as a model
of  compact ring (linear) polymer
which fills the lattice completely \cite{ClJa:polymerE}.
It is also relevant to the protein folding  problem
\cite{ChDi:protein,CaTh:minimum,DoGaOr:proteinfolding}.

Among the properties of Hamiltonian cycles,
the number of Hamiltonian cycles on a given graph is one of the most
fundamental and interesting quantities.  The number is directly
related to the entropy of a lattice polymer in the compact phase. 
When it is non-zero, it also coincides with the degeneracy of optimal
solutions to the traveling salesman problem on the graph.  The number has
been calculated exactly or numerically for a number of fixed regular
lattices
\cite{Kasteleyn,Lieb,ScHiKl:compact,Suzuki:regular,BaSuYu:honyeycomb,%
BaBlNiYu:packedloop,JaKo:fieldtheory}.

In this article, 
I am concerned with random lattices and study the number of
Hamiltonian cycles on them.
Specifically, 
I exactly evaluate the number 
$F_n^{(g)}:=\sum_{G \in S^{n,g}} \mathcal{H}(G)/(\#\mathrm{Aut}\  G)$,
where the number of Hamiltonian cycles on a given graph $G$ is denoted 
by $\mathcal{H}(G)$.
The ensemble $S^{n,g}$ is the set of all trivalent fat graphs
that have $n$ vertices and can be drawn faithfully on a surface of
genus $g$ ( but not on that of $g-1$). See Fig. \ref{fig:fatgraph} for
an example. 
The integer $\#\mathrm{Aut}\  G$ is  related to the symmetry of $G$
and is defined precisely in section \ref{definitions}.
This result extends those of refs.
\cite{EyGuKr:hamiltonian,Higuchi:hamiltonianrandom} where only the
planar case ($g=0$) has been studied.  

The case $g=1$ (torus) is especially interesting in that 
it can be compared with the corresponding problem on a two-dimensional
fixed regular lattice with the periodic boundary condition\footnote{
  In some literatures, the loops is said to be a walk satisfying
  the periodic boundary condition. In the present context, however,
  it is imposed on lattices in the absence of loops or walks.
  }  
for each of the two directions \cite{cond-mat/9711152}.
This comparison provides a good illustration of nature of statistical
systems on random lattices. 

{}From the exact integral expression of $F_n^{(g)}$,
one can extract the large-$n$ asymptotics of the `random average'
of $\mathcal{H}(G)$. 
It is found that the site entropy is independent
of the genus $g$ while the conformational exponent $\gamma$
depends on it linearly.
This behavior is consistent with the KPZ-DDK scaling 
\cite{KnPoZa:LightCone,DiKa:Liouville,David:Liouville}
for two-dimensional gravity coupled to $c=-2$ conformal matter
\cite{EyGuKr:hamiltonian,Higuchi:hamiltonianrandom}. 

The organization of the paper is as follows.
I give definitions and fix notations in section \ref{definitions}.
In section \ref{torus}, the calculation for a random graph on a torus is
presented, putting emphasis on the difference from the 
case of a fixed regular lattice on a torus.
The analysis is extended to the case of surfaces of an arbitrary genus
to yield a simple integral expression in section \ref{surface}.
I discuss my results in section \ref{discussions}.

\section{Definitions} \label{definitions}
Definitions in this section generalize those given in ref.
\cite{Higuchi:hamiltonianrandom}.

\begin{figure}[tb]
  \begin{center}
    \leavevmode
\includegraphics[scale=0.4]{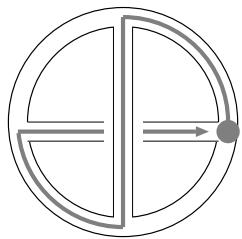}        
\caption{An example of a trivalent fat graph $G\in S^{4,1}$
  which can be drawn on a torus but not on a sphere.
  A Hamiltonian cycle is drawn in a gray line.
  For this graph, $\mathcal{H}(G)=8$ because 
  a direction (arrow) and a base point (dot) are associated with a
  Hamiltonian cycle. 
\label{fig:fatgraph}}
  \end{center}
\end{figure}

Let $S^{n,g}$ be the set of all trivalent fat graphs 
that have  $n$ vertices  possibly with multiple-edges and self-loops
and can be drawn faithfully on an orientable  surface 
$\Sigma_g$ of genus $g\ge0$ (but not on $\Sigma_{g-1}$).  
An example of a trivalent fat graph is drawn in
Fig \ref{fig:fatgraph}. 
Graphs that are isomorphic are identified.
The set $\tilde{S}^{n,g}$ is the labeled version of $S^{n,g}$,
namely, vertices of $\tilde{G}\in\tilde{S}^{n,g}$ are labeled 
$1,\ldots,n$ 
and $\tilde{G}_1, \tilde{G}_2\in\tilde{S}^{n,g}$ are considered identical
only if a graph isomorphism preserves labels.
The symmetric group of degree $n$
naturally acts on $\tilde{S}^{n,g}$ by the label permutation.
The stabilizer subgroup of $\tilde{G}$ is called
the automorphism group $\mathrm{Aut\ }G$.

A Hamiltonian cycle of a labeled graph $\tilde{G}\in \tilde{S}^{n,g}$ is a
directed closed path (consecutive distinct edges connected at vertices) 
which visits every vertex once and only once.
Hamiltonian cycles are understood as furnished with a direction
and a base point (denoted by an arrow and a dot in figures).
The number of Hamiltonian cycles of $\tilde{G}$ is denoted by 
$\mathcal{H}(G)$ because it is independent of the way of labeling.

The quantity I study in this work is 
\begin{equation}
 F_n^{(g)} := \sum_{G\in S^{n,g}} \mathcal{H}(G)\frac{1}{\#\mathrm{Aut\ } G}
 \label{f_n}
\end{equation}
and the generating function
\begin{equation}
  F^{(g)}(p):= \sum_{n=0}^\infty p^n  F_n^{(g)}.
\end{equation}
The key observation made in ref. \cite{Higuchi:hamiltonianrandom}
has been  that $F_n^{(0)}$ can be written as 
the number of isomorphism classes of the pair (graph, Hamiltonian
cycle) and thus is an integer.
It is also true for $F_n^{(g)}$ with $g>0$ :
\begin{equation}
  F_n^{(g)} = \# ( \{ (\tilde{G}, C^n) | 
 \tilde{G}\in \tilde{S}^{n,g}, 
C^n: \text{Hamiltonian cycle on }\tilde{G} \}/\sim),
  \label{num_of_pairs}
\end{equation}
where $(\tilde{G}_1,C^n_1)\sim(\tilde{G}_2,C^n_2)$ 
if and only if
$G_1$ and $G_2$ are isomorphic (forgetting the labels)
and the isomorphism maps $C^n_1$ onto $C^n_2$ 
with the direction and the base point preserved.

\section{Lattices on a torus}\label{torus}
\begin{figure}[htbp]
  \begin{center}
    \leavevmode
\includegraphics[scale=0.7]{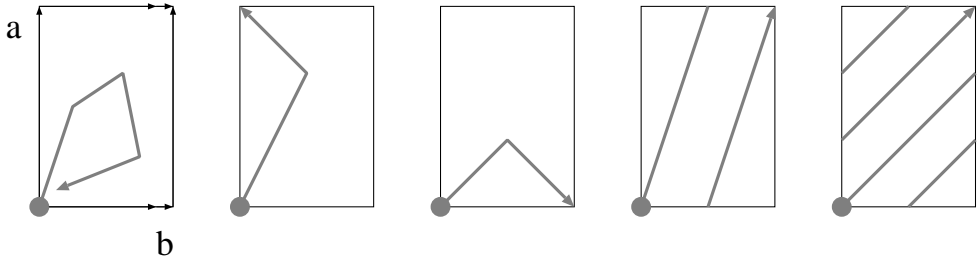}        
\caption{
  Examples of the ways how self-avoiding loops (gray lines) wind around tori.
  A torus $\Sigma_1$ is drawn as a rectangle with two pairs of edges
  identified, with the generator $a,b\in \pi_1(\Sigma_1)$.
  From left to right, the loop represents 
  $0, a,b,2a+b$, and 
  $2a+3b\in \pi_1(\Sigma_1)$, respectively.
    \label{fig:rectangles}}
  \end{center}
\end{figure}

Before proceeding to the calculation of $F^{(1)}_n$,
let us take a glance at the case of fixed regular lattices on a torus
$\Sigma_1$. 
Given a fixed base point, a Hamiltonian cycle represents an element of
the fundamental group $\pi(\Sigma_1)\simeq \mathbb{Z}a+\mathbb{Z}b$ 
where $a$ and $b$ are the meridian and the longitudinal cycles.
Reflecting this fact, 
there are many topological sectors for self-avoiding loops (thus for
Hamiltonian cycles) as depicted in
Fig \ref{fig:rectangles}.
More precisely, an element $m_1 a + m_2 b\in \pi_1(\Sigma_1)$
can be represented by a self-avoiding loop if and only if the pair
$(m_1,m_2)$ satisfies 
\begin{equation}
(m_1,m_2)=(0,0), (\pm1,0), (0,\pm1), 
\quad\text{or} \quad \mathrm{gcd}[m_1,m_2]=1.
  \label{irreducible_loop}
\end{equation}
This rich structure gives rise to an interesting question: 
how many Hamiltonian cycles belong to a given topological sector ?
But at the same time, 
the existence of many topological modes is an obstacle to 
calculating $\mathcal{H}(G)$.

In the transfer matrix method for calculating $\mathcal{H}(G)$,
one maps the system into a state sum model with a local weight.
Once mapped to a state sum model, 
its transfer matrix can be diagonalized by 
employing the Bethe ansatz 
\cite{Suzuki:regular,BaSuYu:honyeycomb}
or numerical calculation
\cite{BlNi:honeycomb,BaBlNiYu:packedloop}.
In the mapping, it is crucial to avoid contributions from vacuum loops,
or small loops disconnected from the largest loop component in question.
This non-local constraint can be imposed as follows.
First one assigns directions to each loop component and 
makes the weight direction dependent.
The sum over all the assignments is taken.
Then the weight is tuned so as that the cancelation occurs 
between two configurations having the vacuum loop of
the same shape but with the opposite directions.

This procedure is straightforward for lattices with disk topology
because all vacuum loops are topologically trivial.
In the case of cylinder geometry,
vacuum loops winding around the cylinder should be taken care of.
To make the cancelation work, one introduces the `seam' and 
associates additional weights to vacuum loops
going across the seam and winding around the cylinder.

When one further imposes  periodicity to the other direction to have a torus,
one come to have a huge number of topologically inequivalent
vacuum loops as shown in  Fig. \ref{fig:rectangles}.
As far as the present author knows, there is no simple way to avoid
a contribution from each of these vacuum loops in the transfer matrix
method. 
Thus, for fixed lattices on a torus, one has to employ the
direct enumeration or the field theoretic approximation in order to evaluate
$\mathcal{H}(G)$ \cite{cond-mat/9711152}.

Note that the Coulomb gas method
\cite{KodeNi:FPLhoneycomb,JaKo:fieldtheory,KoJa:conformational}, which
is capable of determining exact critical exponents, again works for 
the disk or cylinder topology but not for  the torus topology
\footnote{
In contrast to the case of the Hamiltonian cycles or the compact polymer,
various critical exponents of the dense polymer on torus 
was exactly determined by Duplantier and Saleur \cite{DuSa:exact}.
This is because the dense polymer is universal in the sense that the
exponents does not depend on details of the lattice.
}.

\begin{figure}[tb]
  \begin{center}
    \leavevmode
\includegraphics[scale=1]{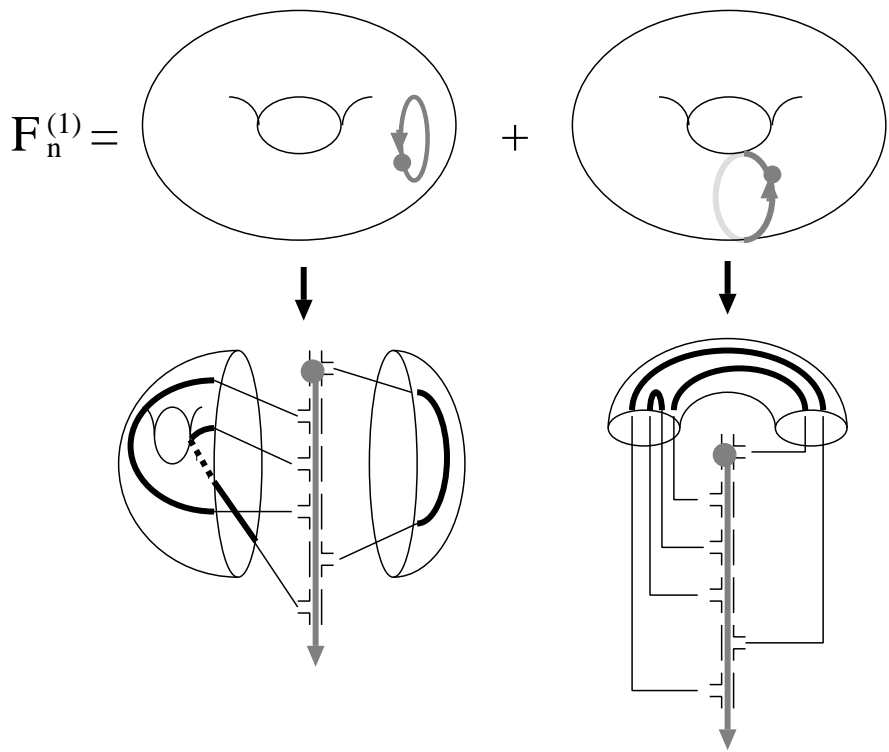}        
\caption{
  Configurations contributing to each of 
  two topological sectors (left: $F^{(1),0}_6$, right: $F^{(1),1}_6$).
  The sequence of T's in the center is obtained by walking along 
  the Hamiltonian cycle (the gray line in the T's).
  The vertical segments of T's on  both sides should be connected by 
  black fat lines on the surface. 
\label{fig:torus}}
  \end{center}
\end{figure}

For random lattices, however, the situation changes dramatically.
Changing the order of the summation, 
I shall sum over random lattice structures first
fixing the topological sector of the Hamiltonian cycle.
Then I sum over the topological sectors.
Because the summation is performed  over random lattices, 
there is no preferred basis $\langle
a,b\rangle$.
One is  free to perform a modular transformation on a torus
% SL$(2,\mathbb{Z})$
to map $m_1 a + m_2 b \mapsto 1\cdot a,  b\mapsto b$ 
if $m_1\neq0$ or $m_2\neq0$.
In other words, only two topological classes of cycles are distinguishable: 
the non-winding sector ($m_1=m_2=0$) 
and  the winding sector ($m_1\neq0$ or $m_2\neq0$).
A contribution from each sector is denoted by
 $F^{(1),0}_n$  and $F^{(1),1}_n$, respectively (Fig \ref{fig:torus})
\begin{equation}
F^{(1)}_n = F^{(1),0}_n + F^{(1),1}_n,\ \ 
F^{(1),w}(p) := \sum_{n=0}^\infty p^n F^{(1),w}_n.
\end{equation}

Imagine that one walks along a Hamiltonian cycle in the specified
direction starting from the base point and records the order of right
and left turns. 
Then one can associate  a sequence of T-shaped objects with the order
as depicted in Fig.\ref{fig:torus}.
If there are $k$ right turns and $n-k$ left ones ($0\le k \le n$), 
there are $\binom{n}{k}$ possible orderings.
The Hamiltonian cycle goes through the horizontal segment of T while 
the vertical segment is left unvisited.
The $n$ vertical segments should be connected pairwisely to 
reproduce the graph completely. 
The pairs are connected with fat black lines 
on the given surface (Fig.\ref{fig:torus}).

Now I consider $F^{(1),0}_n$ and $F^{(1),1}_n$ separately. 

For non-winding sector $F^{(1),0}_n$,
there should be no connection between the right hand side and the left 
hand side of the cycle because it  divides the torus into a disk and 
that with a handle.
Let $A_k^{(g)}$ denote the number of ways of connecting $k$ objects on a disk 
with $g$ handles (but not on that with $g-1$ handles).
I have 
\begin{equation}
F_n^{(1),0}=\sum_{k=0}^n   \binom{n}{k}     
                 A_k^{(0)}     A_{n-k}^{(1)}.
\label{nonwinding}
\end{equation}

On the other hand, for winding sector $F^{(1),1}_n$  I obtain 
\begin{equation}
F_n^{(1),1}=\frac12\sum_{k=0}^n   \binom{n}{k}    
A_{k,n-k}^{(0)},
\label{winding}
\end{equation}
where $A_{k,n-k}^{(g)}$ is  the number of ways of contracting
$k$ and $n-k$ objects on each end of a cylinder with $g$ handles
attached (and with at least one connection between two ends
of the cylinder).

It has been known that $A$'s above can be written as
connected correlation functions of the hermitian gaussian matrix model
\cite{tHooft:largen,BrItPaZu:planar}:
\begin{align}
  \left\langle  \frac{1}{N} \mathrm{tr} M^k \right\rangle
=& \sum_{g=0}^\infty N^{-2g} A_k^{(g)}, 
 \label{tadpole} \\
  \left\langle  \frac{1}{N} \mathrm{tr} M^k \frac{1}{N} \mathrm{tr} M^\ell 
\right\rangle_{\mathrm{c}}
=& \sum_{g=0}^\infty N^{-2g-2} A_{k,\ell}^{(g)}.
  \label{cylinder} 
\end{align}
The expectation value $\langle \cdots\rangle$ is defined by
\begin{equation}
  \langle \cdots \rangle:= \frac{1}{Z} \int dM \ (\cdots)  \ 
e^{-\frac{N}{2}\mathrm{tr}M^2},
\label{expectation}
\end{equation}
where $M$ is an $N\times N$ hermitian matrix variable and the
normalization $Z$ is fixed so as to satisfy  $\langle 1 \rangle =1$.
The subscript c in  $\langle \cdots \rangle_{\mathrm{c}}$ means 
the connected correlation function 
\begin{equation}
 \langle \mathrm{tr} f(M) \ \mathrm{tr} g(M)\rangle_{\mathrm{c}}
:=\langle \mathrm{tr} f(M) \mathrm{tr} g(M)\rangle
-\langle \mathrm{tr} f(M) \rangle \langle \mathrm{tr} g(M)\rangle.
\end{equation}

Because the measure in \eqref{expectation} is gaussian,
$A^{(g)}_k$ and $A^{(g)}_{k,\ell}$ can be 
obtained easily by the method of orthogonal
polynomial \cite{BeItZu:qfttech,Mehta:random2} or by the loop 
equation \cite{Kazakov:kazakovseries,AmChKrMa:beyond}. 
It can be shown that 
\begin{align}
  A^{(0)}(x):=&\sum_{k=0}^\infty x^{-k-1} A^{(0)}_k 
=\frac12\left(p-\sqrt{x^2-4}\right),\\
  A^{(1)}(x):=&\sum_{k=0}^\infty x^{-k-1} A^{(1)}_k =(x^2-4)^{-5/2},\\
  A^{(0)}(x,y):=& \sum_{k,\ell=0}^\infty x^{-k-1} y^{-\ell-1}
  A_{k,\ell}^{(0)}\nonumber \\
=\frac{1}{2^4}
\left( 1 - \frac{x}{\sqrt{x^2-4}}\right)&
\left( 1 - \frac{y}{\sqrt{y^2-4}}\right)
\left( 1 + \frac{\sqrt{x^2-4}-\sqrt{y^2-4}}{x-y}\right)^2.%
\end{align}
Combining these expressions, I finally arrive at exact integral
expressions for  $F^{(1),w}(p)$:
\begin{align}
  F^{(1),0}(p)
=& \oint \frac{dx}{2\pi i} \oint \frac{dy}{2\pi i}
      A^{(0)}(x)      A^{(1)}(y) \frac{1}{1-p(x+y)},\\
  F^{(1),1}(p)
=&\frac12 \oint \frac{dx}{2\pi i} \oint \frac{dy}{2\pi i}
  A^{(0)}(x,y) \frac{1}{1-p(x+y)},
\end{align}
where the contours for $x,y$ go  around the cut $[-2,2]$ counterclockwise.
Careful inspection of the singularities shows that 
\begin{align}
F^{(1),0}(p)
=&\frac{1}{2^{6}\cdot 3 \cdot \pi}(\tfrac14-p)^{-1} 
+ (\text{less singular terms}),\\
F^{(1),1}(p)
=&\frac{1}{2^{7} \cdot \pi}(\tfrac14-p)^{-1} 
+ (\text{less singular terms}).
\end{align}
This implies that $F_n^{(1),w}$ grows as $4^n n^{0}$ for $n\rightarrow
\infty$ 
and that the winding and the non-winding sectors 
contribute with the ratio of $3:2$.

The combinatorial argument leading to \eqref{nonwinding} and
\eqref{winding} was first used by  
Duplantier and Kostov 
in analyzing the dense and dilute phases of polymers on a
\textsl{planar} random lattice \cite{DuKo:spectra,DuKo:random}.
Then it has been explicitly recognized that it can be
equally applied to the compact polymer, or the Hamiltonian cycle problem
on a planar random lattice
\cite{EyGuKr:hamiltonian,Higuchi:hamiltonianrandom}. 
This the first time that the combinatorial argument is successfully
applied to higher-genus cases.

\section{Lattices on surfaces of arbitrary genus}\label{surface}
For random lattices drawn on a surface $\Sigma_g$ of genus $g$,
I find
\begin{equation}
  F_n^{(g)}=
\frac12 \sum_{h=0}^g \sum_{k=0}^n \binom{n}{k}  A_k^{(h)} A_{n-k}^{(g-h)}
+
\frac12              \sum_{k=0}^n \binom{n}{k}  A_{k,n-k}^{(g-1)}
 \label{sectors_arbitrary}
\end{equation}
by enumerating all the topological configurations.
\begin{figure}[tb]
  \begin{center}
    \leavevmode
\includegraphics[scale=0.8]{surface.eps}        
\caption{
  Topologically inequivalent ways to wind a Hamiltonian cycle around a 
  surface of genus $g$.
\label{fig:surface}}
  \end{center}
\end{figure}
Equation \eqref{sectors_arbitrary} can be understood as follows 
(See Fig.\ref{fig:surface}).
When one cuts the surface $\Sigma_g$ along a Hamiltonian cycle,
one is left with a (possibly disconnected) surface with two circular
boundaries.
If the surface remains connected,
the surface obtained should be $\Sigma_{g-1}$ with two holes 
(the second term in Fig \ref{fig:surface}).
If it becomes disconnected, one has $\Sigma_h$ and $\Sigma_{g-h}$
which are to be glued together along the Hamiltonian cycle to form
$\Sigma_g$
(the first term in Fig \ref{fig:surface}).
This exhausts all the possibilities.
There are precisely  $[\tfrac{g}{2}+2]$ topologically inequivalent sectors.
Note that the summands for $h=h_0$ and $h=g-h_0$ in the first term in
\eqref{sectors_arbitrary} 
belong to an identical topological sector.
Equation \eqref{sectors_arbitrary} can be used to calculate
contributions from each topological sectors as well as the sum $F^{(g)}_n$
using \eqref{tadpole} and \eqref{cylinder}.

At this point, I am  naturally led to define the `all-genus'
generating function 
\begin{equation}
  F(p,t)=\sum_{g=0}^\infty t^g \sum_{n=0}^\infty p^n F_n^{(g)}.
\end{equation}
Notably, this simplifies the expression a lot. I find that 
\begin{equation}
  F(p,\tfrac{1}{N^2})=
\frac12\sum_{n=0}^\infty p^n \sum_{k=0}^n \binom{n}{k} 
  \left\langle 
\frac1N \mathrm{tr} M^k\frac1N \mathrm{tr} M^{n-k}
\right\rangle,
  \label{grand}
\end{equation}
where $N$ is the dimension of the matrix variable $M$.
The correlation function in the right hand side
includes both connected and disconnected parts.
In fact, I can easily recover \eqref{sectors_arbitrary} 
by expanding \eqref{grand} in $1/N$.

I further rewrite \eqref{grand} to have a simple integral expression.
I obtain 
\begin{align}
  F(p,\tfrac{1}{N^2})=&\frac12\sum_{n=0}^\infty p^n 
\left( 
\frac{\partial}{\partial t_1} +
\frac{\partial}{\partial t_2} 
 \right)^n
\left.             \left\langle 
\frac1N \mathrm{tr} e^{t_1 M}\frac1N \mathrm{tr} e^{t_2 M}
\right\rangle 
\right|_{t_1,t_2=0}  \nonumber \\
=&\frac12 \int^\infty_0 ds \ e^{-s} 
e^{ps\left( 
\frac{\partial}{\partial t_1} +
\frac{\partial}{\partial t_2} 
 \right)}
\left.           \left\langle 
\frac1N \mathrm{tr}\  e^{t_1 M}\frac1N \mathrm{tr}\  e^{t_2 M}
\right\rangle\right|_{t_1,t_2=0}\nonumber\\
=&\frac12 \int^\infty_0 ds \ e^{-s} 
      \left\langle \left(\frac1N \mathrm{tr}\  e^{spM}\right)^2\right\rangle.
\label{grand_integral}
    \end{align}
In the above derivation, I have made use of the integral expression of
the $\Gamma$-function. 
It can be easily checked that the coefficients of  $N^{-2}$ and $N^0$
in \eqref{grand_integral}
exactly agree with eq. \eqref{torus} and the result in 
ref.\cite{Higuchi:hamiltonianrandom}, respectively.

The expression \eqref{grand_integral} is not only
simple but also practically useful. 
It enables one to calculate $F^{(g)}_n$  by the method of orthogonal
polynomials \cite{BeItZu:qfttech,Mehta:random2}.

Because $F(p,\tfrac{1}{N^2})$ has been written  
as a correlation function of the \textsl{gaussian} hermitian matrix
model in \eqref{grand_integral},
the system should fall into the same universality class as the
topological gravity or two-dimensional quantum gravity 
coupled to $c=-2$ matter \cite[Subsection 9.3]{GiMo:lectures}.
Namely, the double scaling behavior
\begin{equation}
F(p,\tfrac{1}{N^2}) \sim (\tfrac14-p)^{2} f( N^2 (\tfrac14-p)^3)
\end{equation}
or equivalently the large-$n$ asymptotics 
\begin{equation}
F^{(g)}_n \sim 4^n n^{3g-3}    \ \ \ (n\rightarrow \infty)
\label{asymptotics}
\end{equation}
is implied.

The central charge $c=-2$ was previously  obtained by Duplantier and Kostov 
in the analysis of the dense phase of polymers on random lattices of
arbitrary genus \cite{DuKo:random}.
They made use of the hermitian O($n$) multi-matrix
model with a non-gaussian interaction, 
where $1/N$-expansion corresponds to the genus expansion
and the $n\rightarrow0$ guarantees  that a Hamiltonian cycle is
connected \cite{EyGuKr:hamiltonian}. 
In the present analysis of the compact polymer, I have written
the generating function 
$F(p,\tfrac{1}{N^2})$, whose 
$1/N$-expansion is again the genus expansion,
in terms of the simplest matrix model: the hermitian gaussian 1-matrix
model without any limiting procedure.

\section{Discussions} \label{discussions}
I have obtained the generating function for
the number of Hamiltonian cycles on surfaces of arbitrary genus.
For genus $g=1$, the contributions from winding and
non-winding sector are determined.
This can be done for an arbitrary genus by eq. \eqref{sectors_arbitrary}.

The limit of large graphs \eqref{asymptotics}, 
is consistent with the assertion that 
the system is in the same universality class as the $c=-2$ quantum gravity
\cite{EyGuKr:hamiltonian,Higuchi:hamiltonianrandom}. 
In the way of calculating the generating function, 
I have made use of microscopic loop amplitudes of the gaussian 
hermitian 1-matrix model.

Many deep connections have been found between 
the gaussian hermitian matrix model and the $c=-2$ quantum gravity
since the first calculation of quantum 
gravity beyond the spherical limit at $c=-2$ by Kostov and
Mehta \cite{KoMe:arbitrarygenus}. 
They have written the free energy at genus $g$ in terms
of correlation functions of the gaussian hermitian matrix model at
order $1/N^{2g}$.
In contrast, $F^{(g)}(p)$ in the present case gets contributions from 
correlators at all orders above $1/N^{2g}$.
The free energy in ref. \cite{KoMe:arbitrarygenus} is the generating
function for the number of maximal trees on graphs, 
while $F^{(g)}(p)$ here generates the number of  Hamiltonian cycles.

It is interesting to compare eq.\eqref{asymptotics} with the 
number of connected trivalent fat graphs with $n$ vertices 
in the absence of Hamiltonian cycles \cite{EyGuKr:hamiltonian}.
The latter behaves as \cite{BrItPaZu:planar,KaKoMi:DT,David:DT}.
\begin{equation}
  \hat{F}_n^{(g)}
:=\sum_{G\in S^{n,g}} 1 \times \frac{1}{\#\mathrm{Aut\ } G}
 \sim ( 2 \cdot 3^{3/4} )^n n^{\frac52 g - \frac72}.
\label{puregravity}
\end{equation}
I introduce the `random average' of $\mathcal{H}(G)$ 
among $G$'s in $S^{n,g}$ by
\begin{equation}
 \langle \mathcal{H}(G) \rangle_{G\in S^{n,g}} := 
  \frac{F_n^{(g)}}{\hat{F}_n^{(g)}} 
\sim (2\cdot3^{-3/4})^n n^{(g+1)/2}
= (0.877383\cdots)^n n^{(g+1)/2}.
 \label{random_average}
\end{equation}
This behavior may be compared to the case of the fixed flat lattice with the 
same coordination number 3, \textsl{i.e.} the hexagonal lattice. 
The number $\mathcal{H}(G)$ for the hexagonal lattice grows as
\cite{Suzuki:regular,BaSuYu:honyeycomb}  
\begin{equation}
  \mathcal{H}(G) \sim (3^{3/4}/2)^n n^{\gamma} 
   \sim (1.13975\cdots)^n n^{\gamma} \quad(n:=\#G),
 \label{fixed}
\end{equation}
where the conformational exponent $\gamma$ is believed to be unity\footnote{
  This definition of $\gamma$ is the standard one in the case of fixed
  lattice.  One associate a base point to each Hamiltonian cycle in
  the present case, yielding an extra factor $n^1$.
}.
Having this behavior of fixed regular lattice in mind, 
it is puzzling that the average
$ \langle \mathcal{H}(G) \rangle_{G\in S^{n,g}}$ 
decreases as $n$ grows.

These totally different behaviors of \eqref{random_average} and \eqref{fixed} 
can be understood as follows.
In the ensemble $S^{n,g}$ for arbitrarily large $n$, 
there are many $G$'s with $\mathcal{H}(G)=0$.
In fact, if a graph admits a Hamiltonian cycle, then the graph should 
be $2$-edge-connected.
Thus, as pointed out in ref. \cite{EyGuKr:hamiltonian}, 
$G$ does not admit any Hamiltonian cycles if $G$ is one-particle reducible.
Therefore it is more natural to consider the random average over 
$\bar{S}^{n,g}$, the restriction of $S^{n,g}$ to 
one-particle irreducible graphs\footnote{
  In this respect, I have already excluded disconnected
  graphs, which do not admit Hamiltonian cycles, 
 in the definition of the  random average \protect\eqref{random_average}.
}.
It is known  that  the number of one-particle irreducible graphs grows 
as \cite{BrItPaZu:planar}
\begin{equation}
  \bar{F}_n^{(g)}
:=\sum_{G\in \bar{S}^{n,g}} 1 \times \frac{1}{\#\mathrm{Aut\ } G}
 \sim ( 2^{-1/2} \cdot 3^{3/2} )^n n^{\frac52 g - \frac72}.
\end{equation}
Therefore I obtain 
\begin{equation}
\!\!\!\!\!
 \langle \mathcal{H}(G) \rangle_{G\in \bar{S}^{n,g}} := 
  \frac{F_n^{(g)}}{\bar{F}_n^{(g)}} 
\sim (2^{5/2}\cdot3^{-3/2})^n n^{(g+1)/2}
= (1.08866\cdots)^n n^{(g+1)/2}.
\label{random_average_1pi}
\end{equation}

It is interesting to note that the value $1.08866\cdots$ is 
near to  the corresponding value $1.13975\cdots$ 
for the fixed hexagonal lattice as well as 
the field theoretic estimate $3/e=1.10364\cdots$ for regular lattices with
coordination number three \cite{OrItDo:hamiltonian,cond-mat/9711152}.
For $g=1$, 
the power correction in \eqref{random_average_1pi} becomes $n^1$,
which coincides with that of the fixed flat hexagonal lattice \eqref{fixed}.
One is tempted to speculate that 
Hamiltonian cycles on random lattices resembles 
those on a regular lattice 
with the identical coordination number and the identical average curvature 
(zero for $g=1$) once one restricts oneself to one-particle irreducible graphs.

In ref. \cite{EyGuKr:hamiltonian}, Eynard, Guitter and Kristjansen 
have mapped the problem of counting Hamiltonian cycles on planar random
lattices to that of O($n$) model in the  $n\rightarrow0$ limit on the sphere.
Their Hamiltonian cycles are not furnished with directions and base points.
The $1/N$-expansion of their model should contain the information of
the number of Hamiltonian cycles on surfaces of arbitrary genus.
I believe that their approach and the present one are complementary
each other.

Equation \eqref{grand_integral} formally resembles the expression for
macroscopic loop amplitudes of the  one-matrix model in
ref. \cite{BaDoSeSh:loops}. 
It may be interesting to translate the present analysis into the
language of free fermions.
\medskip

\subsection*{Acknowledgments}
I thank Shinobu Hikami and Mitsuhiro Kato for useful discussions.
This work was supported by the Ministry of Education, Science
and Culture under Grant 08454106 and 10740108 
and by Japan Science and Technology Corporation under CREST.  \medskip

\bibliographystyle{physlett}
\bibliography{shrtjour,mrabbrev,polymer,2dgrav,matrix,mypubl}

\end{document}